\documentclass[12pt,preprint]{aastex}

\shorttitle{Modelling Stellar Pollution}
\shortauthors{Aaron Dotter and Brian Chaboyer}

\begin{document}

\title{The Impact of Pollution on Stellar Evolution Models}
\author{Aaron Dotter and Brian Chaboyer}
\affil{Department of Physics and Astronomy, Dartmouth College, 6127 Wilder Laboratory, Hanover, 
NH 03755}
\email{Brian.Chaboyer@Dartmouth.edu}

\begin{abstract}
An approach is introduced for incorporating the concept of stellar
pollution into stellar evolution models. The approach involves
enhancing the metal content of the surface layers of stellar
models. In addition, the surface layers of stars in the mass range of
$0.5-2.0\,\mathrm{M}_{\odot}$ are mixed to an artificial depth
motivated by observations of lithium abundance. The behavior of
polluted stellar evolution models is explored assuming the pollution
occurs after the star has left the fully convective pre main sequence
phase. Stellar models polluted with a few Earth masses
($\mathrm{M}_{\oplus}$) of iron are significantly hotter than stars of
the same mass with an equivalent bulk metallicity. Polluted stellar
evolution models can successfully reproduce the metal-rich, parent
star $\tau$ Bootis and suggest a slightly lower mass than standard
evolution models. Finally, the possibility that stars in the Hyades
open cluster have accreted an average of $0.5\,\mathrm{M}_{\oplus}$ of
iron is explored. The results indicate that it is not possible to rule
out stellar pollution on this scale from the scatter of Hyades stars
on a color-magnitude diagram. The small amount of scatter in the
observational data set does rule out pollution on the order of
$\sim1.5\,\mathrm{M}_{\oplus}$ of iron. Pollution effects at the low
level of $0.5\,\mathrm{M}_{\oplus}$ of iron do not produce substantial
changes in a star's evolution.
\end{abstract}


\section{Introduction}
The discoveries of dozens of giant planets in orbits with semi-major
axes less than 1 AU (for a complete listing see www.exoplanets.org),
which began in 1995 \citep{mq}, posed a problem for planet formation
theories. Such theories indicate that giant planets ought to form in
orbits with semi-major axes greater than 1 AU.  However, through
interactions with other planets and the protoplanetary disk it is
possible for these planets to migrate inward until they reach a stable
orbit \citep{lin}. The inward migration can sweep material in the disk
onto the surface of the star. \citet{gonz2} pointed out that if the
accreted material is metal-rich then stars with planets (SWPs) will
exhibit enhanced metallicity, a concept referred to as ``stellar
pollution''.

The case has been made in recent years that SWPs are metal rich,
for example  \citet{laws}. The two main explanations for these stars to be 
metal-rich are a high intrinsic metallicity and the accretion of $\sim$5 
$\mathrm{M}_{\oplus}$ of iron from the protoplanetary disk, for a more 
complete discussion see \citet{mc}. 

Based upon the apparent deficiency of heavy elements in the asteroid
belt, and considerations of orbital dynamics in the early solar
system, \citet{mea} found that the formation of Jupiter likely led to
the accretion of $\sim$0.5 $\mathrm{M}_{\oplus}$ of iron onto the Sun. Their analysis
of the correlations between [Fe/H]--age--mass in local field stars
lead \citet{mea} to suggest that stellar pollution effects in the
solar neighborhood appear to be quite common at a similar level.
Using a Monte Carlo approach they estimate that field stars have
accreted an average of $\sim$0.5 $\mathrm{M}_{\oplus}$ of iron with standard
deviation equal to half the mean.  Hence, \citet{mea} suggest that 
most protoplanetary disks are similar to that of the Sun and not the 
stars in the SWPs sample.

In a paper on modeling the evolution of SWPs, \citet{ford} discuss the need 
for stellar evolution models which have enhanced surface metallicities 
relative to the interiors. The authors briefly explain their approach to the 
problem which involves increasing the metallicity of the surface convection
zone and/or decreasing the interior metallicity, and discuss the 
impact of such models on the Sun and a few SWPs. The main thrust of
the paper is, however, modeling SWPs with standard stellar evolution 
models. 

In this paper, stellar evolution models are presented which allow for
the addition of metal-rich material to the surface. The behavior of
these models is compared standard stellar evolution models. The
modifications made to standard stellar evolution models are explained
in $\S$2. The basic impact of these changes on the evolution of a star
is presented $\S$3. In $\S$4 the metal-rich parent star $\tau$ Bootis
is modeled with both polluted and unpolluted models and the results
are compared. In $\S$5 polluted model isochrones are compared to
observational data on the Hyades; the likelihood that the Hyades stars
have experienced low-level pollution is discussed. The attempt to
create polluted stellar evolution models is summarized in $\S$6.

\section{Stellar Models}
Stellar evolution models were constructed using the Dartmouth stellar 
evolution code \citep{cea}. 
Helium and heavy element diffusion are included using \citet{thoul}. All
models use a solar calibrated mixing length $\alpha$ = 1.75.
The models were evolved from pre main sequence, fully convective polytropes to
the main sequence turnoff in $\sim$800 time steps and include overshooting of
the surface convection zone.


\citet{mea} suggest that we may model the depletion 
of lithium from the photosphere by introducing a convective overshoot 
to the surface convection zone. They present 
a model for the mass of the surface mixed layer (SML) as a function of stellar
mass and [Fe/H] which has been implemented in the code for this work. Each 
time the code calls for the surface convection zone to be 
mixed we calculate the mass of the SML and mix down to its base. 
For a plot of SML mass as a function of stellar mass, see Figure 2 of 
\citet{mea}. 

In the literature on stellar pollution the amount of accreted material is 
often referred to in Earth masses ($\mathrm{M}_{\oplus}$) of iron. We shall follow this
 convention throughout. Following arguments made by \citet{mea}, protoplanetary
material is roughly 20\% iron by mass. Note that
we are essentially changing the mass fraction of heavy elements (Z) in the 
models and the accretion of 1 $\mathrm{M}_{\oplus}$ of iron corresponds to the addition
of 5 $\mathrm{M}_{\oplus}$ of Z.

Stars which accret rocky material will experience an increase in heavy 
element composition in the outer layers. Our approach is to rescale the 
composition of the SML without altering its mass. Starting with an amount of 
iron, we take the ratio of the mass of rocky material to the mass of the SML 
($\mathrm{Z_{rock}}$) and 
add this to the initial mass fraction of metals ($\mathrm{Z_i}$). The new mass fraction
of metals is 
\begin{equation}
Z_f = Z_i + Z_{rock}
\end{equation}

A new heavy element mass fraction is calculated and the abundances of hydrogen
 (X) and helium (Y) are adjusted. We rescale X and Y as follows
\begin{equation}
\frac{X_f}{X_i} = 1 - \frac{Z_{rock}}{1-Z_i}
\end{equation}
\begin{equation}
\frac{Y_f}{Y_i} = 1 - \frac{Z_{rock}}{1-Z_i}
\end{equation}
The $i$- and $f$-subscripts refer to initial and final compositions,
respectively.

In models which accrete more than about $1\,\mathrm{M}_\oplus$ of iron we found
it necessary to increase the metallicity over multiple time
steps. Spreading the pollution over several time steps allows us to
avoid the problems which occur when the conditions in the model change
too quickly for the code to adjust. In addition, it is more in line
with current theories which describe the time scales of stellar
pollution as taking place over tens of millions of years. Neither
altering the time scale nor changing the exact age at which rocky
material is accreted (after the zero-age main sequence, ZAMS) has a
significant impact on evolution.  Unless otherwise noted, we begin to
add the rocky material when the model reaches the ZAMS\footnote{
We note that if the pollution
were to occur significantly before the the star reaches the ZAMS, then
accreted material would be mixed throughout the star, as stars in the
early pre main sequence phase of evolution are fully
convective. For stellar pollution to have observational
consequences, it must occur after the star's convection zone has
thinned near the end of the pre main sequence.
}. Since all stars considered metal rich compared to the Sun we rescale 
Y and Z (from Solar values) in the ratio $\Delta$Y/$\Delta$Z = 2 
\citep{cgl}.  

Finally, to calculate [Fe/H] it is sufficient to use [Fe/H]=log($mathrm{Z_*/Z_{\odot}}$) since the relative change in X is insignificant compared to the relative 
change in Z. However, this assumes the star in question has
the same abundance of iron as the Sun. The accreted material is
roughly 20\% iron by mass, while the Sun is only about 0.14\% iron by
mass (iron comprises 7.3\% of $\mathrm{Z_{\odot}}$ by mass, and
$\mathrm{Z_{\odot}}$=0.019). We account for this difference by introducing a
factor to the [Fe/H] formula which accounts for the increased
abundance of iron on the surface of the star.
\begin{equation}
\left[\frac{Fe}{H}\right]=log\left(\frac{Z_*}{Z_{\odot}}\right)+log\left(\frac{M_P(Fe) + M_{SML}(Fe)}{M_{SML}(Fe)}\right)
\end{equation}
Where $\mathrm{M_P}$(Fe) is the mass of iron in the polluting material and 
$\mathrm{M_{SML}}$(Fe) is the mass of iron in the SML.



\section{Basic Results}
The most significant impact of stellar pollution on the evolution of star
is the difference in effective temperature. Given two main sequence stars of 
similar mass,
age, and surface metallicity a polluted star will have a higher effective 
temperature than an unpolluted star (for a specific example see $\S$4). 

The altered behavior can be interpreted as follows: the unpolluted
model has a higher mass fraction of heavy elements in the deep interior
(below the SML).  This leads to a higher opacity in the deep interior
of the unpolluted model, making it more difficult for energy to
flow outward.  The unpolluted model is thus physically larger and 
cooler than the polluted model.

The effect of adding a given amount of iron to a star decreases with SML mass 
because mixing the polluting material over a larger region dilutes the effect.

\section{$\tau$ Bootis}

Recent evidence that SWPs are metal-rich \citep{laws} compared to a
sample of field stars lends support to the concept that these stars
have accreted metal-rich material before the planetary system reached
equilibrium. If the accretion occurred after the star left the
pre main sequence convective phase the effects should be noticeable in
the form of high metallicity.

$\tau$ Bootis has a short-period, giant planet. It also has a high
metallicity, [Fe/H]=0.32$\pm$0.06 \citep{gonz}. To illustrate the
effects stellar pollution may have when stellar models are used to infer
properties of real stars, observational data for $\tau$ Boo was fitted
by standard and polluted models.

\begin{figure}
\plotone{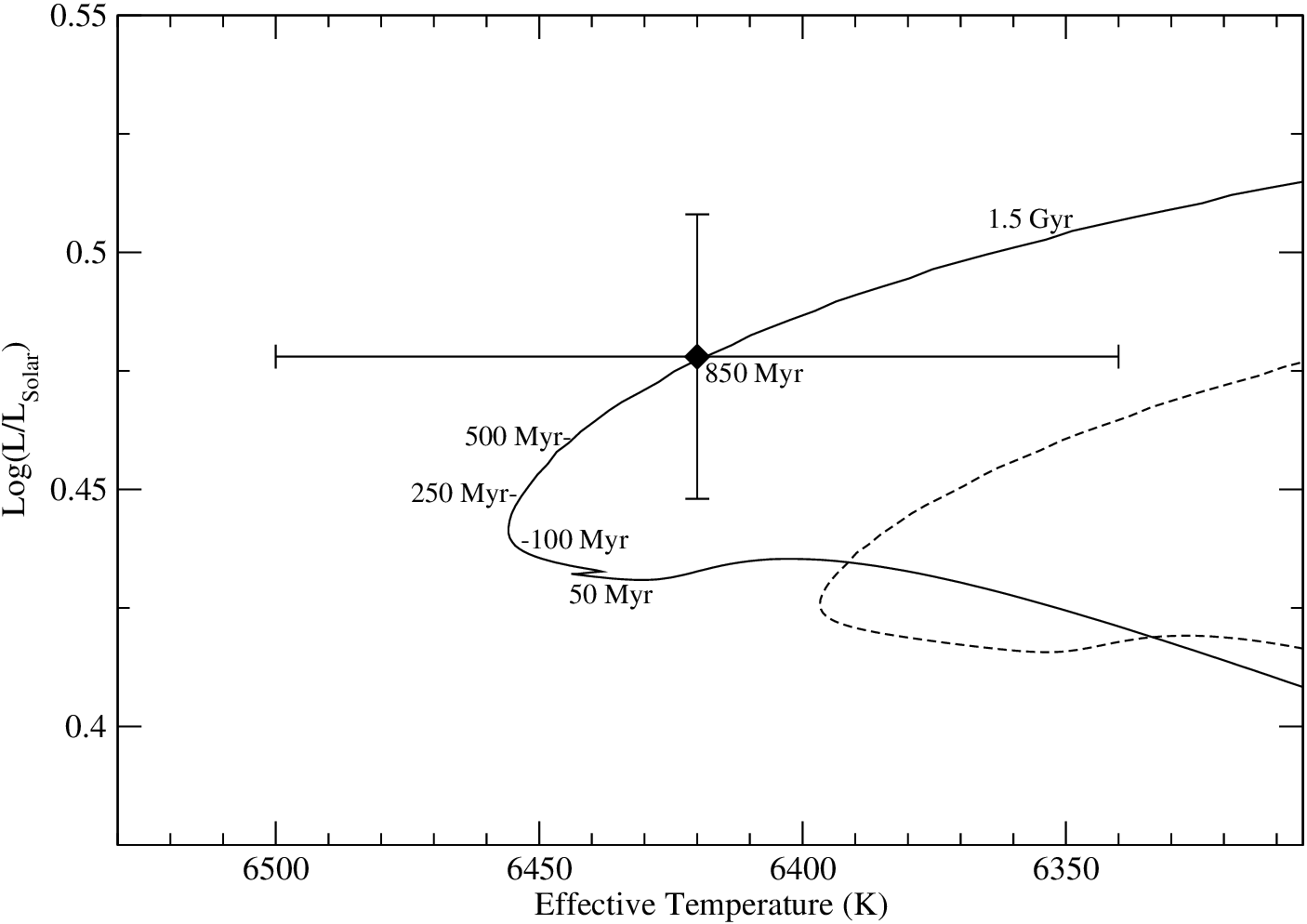}
\caption{H-R diagram showing polluted (solid line) and unpolluted (dashed line) models with M = 1.36 $\mathrm{M}_{\odot}$. Both models have the same [Fe/H] at the surface. The error bars represent observationally derived values for $\tau$ Boo. The polluted model received 2 $\mathrm{M}_{\oplus}$ of iron at the ZAMS.\label{tbHR}}
\end{figure}

A series of polluted models was constructed with masses between 1.30 and 1.40 
$\mathrm{M}_{\odot}$ 
adding 1-5 $\mathrm{M}_{\oplus}$ of iron when the model reaches the ZAMS. 
Figure \ref{tbHR} shows the evolutionary tracks of two 1.36 $\mathrm{M}_{\odot}$ 
stellar models. The solid line is a model polluted with 2 $\mathrm{M}_{\oplus}$ of iron.
The dashed line is an unpolluted model with bulk metallicity chosen to match
the surface metallicity of the polluted model. The polluted model is marked at several points to show how
the model evolves with time. The unpolluted model evolves in the same manner.

\begin{deluxetable}{ccccccc}
\tablecolumns{7} 
\tablewidth{0pc}  
\tablecaption{$\tau$ Boo Model Results\label{tbtable}} 
\tablehead{ 
\colhead{Mass ($\mathrm{M}_{\odot}$)} & \colhead{Log(L/$\mathrm{L_{\odot}}$)}   & \colhead{$\mathrm{T_{eff}}$ (K)}
& \colhead{[Fe/H]} 
& \colhead{$\mathrm{Z_{init}}$}   & \colhead{$\mathrm{Z_{rock}}$ ($\mathrm{M}_{\oplus}$ Fe)}    
& \colhead{Age (Gyr)}}
\startdata

\sidehead{Observational Data}
\nodata & 0.478${+0.030\atop -0.032}$\tablenotemark{a} 
& 6420$\pm$80\tablenotemark{b} & 0.32$\pm$0.06\tablenotemark{b} 
& \nodata & \nodata & \nodata \\

\sidehead{Polluted Models}
1.380 & 0.486 & 6435 & 0.364 & 0.039 & 1  & 0.66 \\
1.360 & 0.478 & 6420 & 0.362 & 0.035 & 2 & 0.85 \\ 
1.345 & 0.500 & 6435 & 0.345 & 0.030 & 3 & 1.18  \\
1.330 & 0.476 & 6415 & 0.329 & 0.028 & 3 & 1.20    \\
1.320 & 0.478 & 6420 & 0.340 & 0.025 & 4 & 1.35     \\
1.300 & 0.478 & 6420 & 0.325 & 0.023 & 4 & 1.54      \\

\sidehead{Unpolluted Models} 
1.400 & 0.497 & 6450 & 0.375 & 0.043 & \nodata& 0.51   \\
1.380 & 0.478 & 6420 & 0.335 & 0.040 & \nodata & 0.64  \\ 
1.360 & 0.479 & 6430 & 0.278 & 0.035 & \nodata & 0.85    \\
1.345 & 0.490 & 6435 & 0.275 & 0.035 & \nodata & 0.98    \\

\enddata 
\tablenotetext{a}{\citet{ford}}
\tablenotetext{b}{\citet{gonz}}
\end{deluxetable} 

Table \ref{tbtable} contains a small sample of polluted models with 
luminosities, temperatures, and ages at closest approach to the observational 
data.
 For comparison we include models constructed using the same evolution code but
which have not been polluted.

The data in Table \ref{tbtable} show that, for a given mass,  the
polluted models are slightly older than the unpolluted models: from a
few percent at the high mass end (around 1.40 $\mathrm{M}_{\odot}$) up to 20\%
at the low mass end (1.34 $\mathrm{M}_{\odot}$).  The polluted
models have a best fit mass which is 0.02 - 0.03 $\mathrm{M}_{\odot}$ lower than the
best fit from the unpolluted models. The purpose of this table is only to 
illustrate the general trends which pollution has on stellar models, and 
not to constrain the stellar mass of, or amount of 
pollution, experienced by, $\tau$ Boo.

\section{The Hyades}

\citet{mea} estimate that stars in the solar neighborhood have accreted 
0.5$\pm$0.25 $\mathrm{M}_{\oplus}$ of iron. The sample used to make this
estimate was comprised of field stars which have a broad range of ages
and bulk metallicities.  The variation in age and metallicity of the
sample, along with the fact that younger stars tend to be more
metal-rich than older stars, complicated the analysis of \citet{mea}.

Stars in the Hyades are likely to have formed at the same time from
material of the same composition. As a result, Hyades stars are likely
to yield better constraints than field stars on stellar pollution.
The relatively young age of stars in the Hyades, combined with a
uniform composition, should imply a tight main sequence. However, planet
formation and accretion of metal-rich material will vary from one star
to another which will cause scatter in the color-magnitude diagram (CMD). 

\citet{acq} proposes that the scatter of Hyades stars about a theoretical
isochrone can be used to limit the level of pollution experienced by Hyades
stars.  \citet{acq} did not
have access to polluted stellar evolution models, and so based her
proposal on some plausible assumptions that the effects of stellar
pollution would have on the evolution of stars. With the modified evolution
code presented in this paper it is possible to directly test Quillen's 
proposal.

As a preliminary test, three isochrones were constructed with varying
levels of pollution. Each isochrone consists of models ranging in mass
from 0.6 to 1.8 $\mathrm{M}_{\odot}$ in steps of 0.05
$\mathrm{M}_{\odot}$.  Each model begins on the pre main sequence,
receives the polluting material at about 150 Myr (at which point all
models have reached or passed the ZAMS), and is evolved to 650
Myr. Uncertainty in the final age of the isochrones by $\pm$50 Myr
alters the scatter by an amount roughly an order of magnitude below
the scatter due to pollution effects.

 The models in each isochrone have a common initial $Z=0.024$ leading
to $<$[Fe/H]$>\,\simeq\,$0.13 in accord with \citet{friel} who find
$<$[Fe/H]$>$ = 0.127$\pm$0.022 in the range of 6000 - 7000 K. In the
``mean'' polluted isochrone, models are polluted with 0.5
$\mathrm{M}_{\oplus}$ of iron. The other two isochrones correspond to
pollution amounts of $2\,\sigma$ above and below the mean,
respectively 1.0 and 0.0 $\mathrm{M}_{\oplus}$ of iron.

\begin{figure}
\plotone{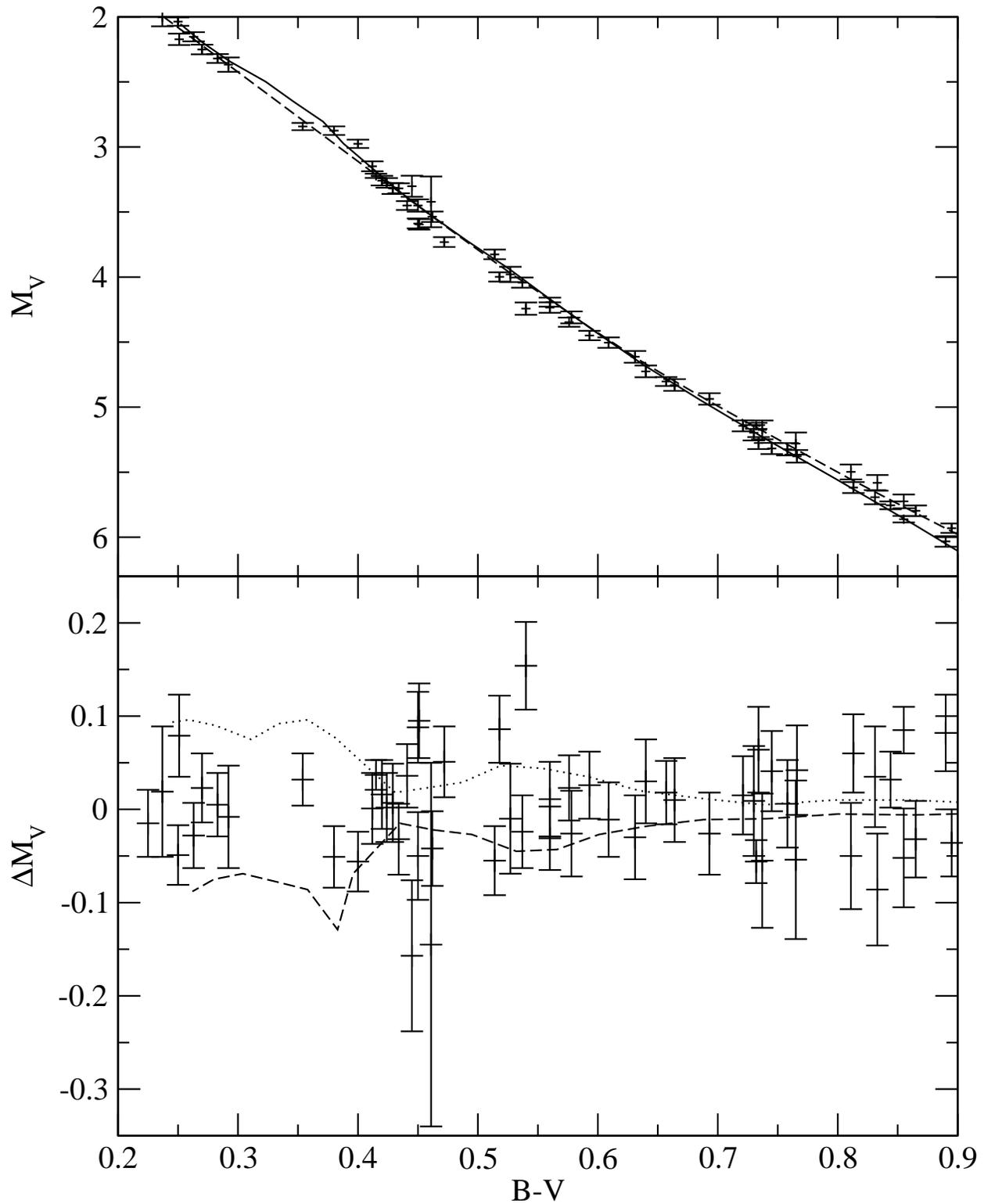}
\caption{$Top\,Panel$--A CMD of the Hyades sample with the mean polluted 
isochrone (solid line) and LOWESS fit (dashed line) superimposed. $Bottom\,Panel$--A scatter plot in $\Delta \mathrm{M_V}$ of the
Hyades sample about the LOWESS fit. Superimposed are the
differences between the $2\,\sigma$ isochrones and the mean
isochrone. (dashed line: $+2\,\sigma$; dotted line: $-2\,\sigma$).\label{hyaHR}}
\end{figure}

The top panel in Figure \ref{hyaHR} is a CMD showing the main sequence
from 0.6 -- 1.8 $\mathrm{M}_{\odot}$ of the Hipparcos Hyades sample of
\citet{deB} (also used by \citet{acq} ).  \citet{kur} color tables
were used to convert the models from luminosity and effective
temperature to $\mathrm{M_V}$ and B--V.  In addition, the CMD plot
includes the mean polluted isochrone and a fit to the data made using
the LOWESS (Locally Weighted Regression) technique
\citep{cleve,geb1,geb2}.  In the plot the dashed line is the LOWESS
fit and the solid line is the mean isochrone.  The LOWESS fit is
weighted by the the local points, and represents the best fit to the
data that is possible with a single line.

The bottom panel of Figure \ref{hyaHR} shows the scatter of the Hyades
sample about the LOWESS fit. Superimposed are the differences between
the $2\,\sigma$ isochrones and the mean isochrone. The difference between the
mean and $+2\,\sigma$ isochrone is represented by the dashed line; the
difference between the mean and the $-2\,\sigma$ isochrone is
represented by the dotted line.  Note that the dispersion of the data
points is smaller than that predicted by the $\pm 2\,\sigma$ models,
showing that the dispersion predicted by the stellar pollution models is
not ruled out by the data.

The LOWESS technique is used to fit the Hipparcos Hyades sample
because the mean polluted isochrone diverges slightly from the main
sequence for $\mathrm{B-V}$ greater than $\sim$0.65 mag. This
difference between the models and the data is likely due to problems
with the treatment of convection, or in the color calibration of the
models.  We are not interested in studying this problem in this paper, rather 
the goal is to compare the scatter in the Hyades stars about a smooth main
sequence to the predicted scatter caused by pollution. The LOWESS fit allows
a comparison between the observed scatter and the theoretical prediction 
which is not hampered by large scale errors introduced by the color conversion and/or treatment of convection.  In general, the theoretical isochrone is a reasonable fit to the data, with the the only consistent deviation between the 
LOWESS fit and the mean polluted isochrone occurring for $\mathrm{B-V}\ge0.65$ 
mag.

The fit of the mean polluted isochrone to the observational data has a
reduced $\chi^2$ of $\sim$3.3 and the LOWESS fit has a reduced
$\chi^2$ of 2.1. The mean polluted isochrone has the lowest $\chi^2$
of all the isochrones constructed and is therefore the standard. The
LOWESS fit is the best is possible single line fit to the data, and
the large value of $\chi^2$ implies that their is an inherent
dispersion in the data.

As a more stringent test, the method of \citet{geb1,geb2} for
determining the velocity dispersion as a function of radius within a
globular cluster has been adapted to study the scatter in the CMD.
In this method, the value of the error bar for each star in
 the Hyades sample is removed in quadrature from the absolute value of 
$\Delta \mathrm{M_V}$ 
(scatter) value of that star. 
The square root of the resulting value is the $\mathrm{M_V}$ dispersion, 
analogous to 
the velocity dispersion in the globular cluster studies of \citet{geb1,geb2}. 
If the size of the error
bar is larger than the absolute value of $\Delta \mathrm{M_V}$ for a given data
point, the $\mathrm{M_V}$ dispersion is set to zero for that point. The LOWESS 
fit is then determined for $\Delta \mathrm{M_V}$ as a function of \bv.
This produces ``dispersion profile'' for the Hyades sample. 
Note that the dispersion is a non-negative quantity, any information about the
scatter being above or below the mean isochrone is lost.
 
The final step is to estimate the error in the derived dispersion
profile.  This is done in a manner somewhat analogous to the standard
bootstrap re-sampling technique for error analysis.  In particular, for
each data point one randomly selects a new value for $\mathrm{M_V}$
using a Gaussian distribution, with a mean equal to the original 
value of $\mathrm{M_V}$ and a one-$\sigma$ width equal to the 
quoted one-sigma error in the data point.  The process is performed one
thousand times for each star and a mean dispersion profile with
confidence bands is calculated from the results.

\begin{figure}
\plotone{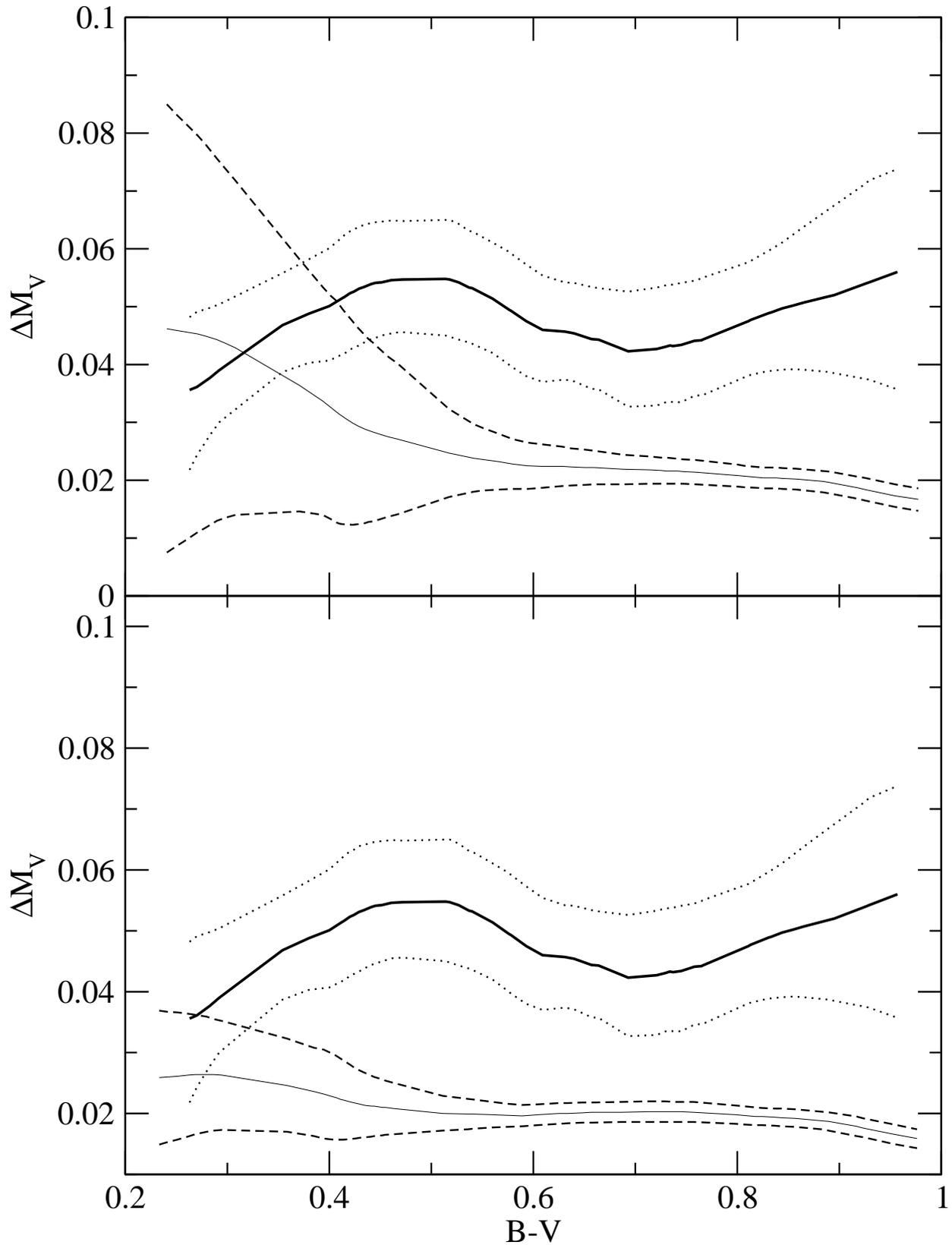}
\caption{$Top\,Panel$--Comparison of the dispersion in the Hyades sample 
(thick line) and the average of the mock data sets with mean pollution of 1.0 
$\mathrm{M}_{\oplus}$ (thin line). 90\% confidence bands are drawn for the 
dispersion in the data (dotted lines) and the mock data sets (dashed lines). 
$Bottom\,Panel$--Same as the top panel but for the 0.5 $\mathrm{M}_{\oplus}$
 case.\label{disp}}
\end{figure}

Having obtained the dispersion profile for the Hyades data, it is
necessary to put the theoretical prediction in a similar form for
comparison. Using the mean polluted isochrone as a guide, the mass of
each star with $0.2\le \mathrm{B-V}\le 1.0$ in the Hyades sample was
determined. Stellar models with these particular values of mass were
evolved 1000 times, where each of the stellar evolution models had a
random amount of stellar pollution added to it at 150 Myr.  The amount
of the polluting material was drawn from a Gaussian distribution with
standard deviation equal to half the mean. This is done for two cases,
means equal to 0.5 and 1.0 $\mathrm{M}_{\oplus}$ of iron.

The result of this simulation is a set of 2000 theoretical CMDs,
where each CMD contains the same number of data points as the Hyades
data set, but for each each of the stars in the CMD has been polluted
by a varying amount.  To compare with the dispersion profile in the
data, the individual theoretical CMD is treated as a ``mock'' data set
and put through the exact same procedure as the real data set to
generate a dispersion profile of $\mathrm{M_V}$ as a function of \bv.
This was done each of the 2000 theoretical CMDs, first for 1000 CMDs which had a mean amount of pollution equal to 0.5 $\mathrm{M}_{\oplus}$ of iron and then for the 1000 CMDs with a mean pollution of 1.0 $\mathrm{M}_{\oplus}$ of iron.

Figure \ref{disp} is a plot of the dispersion in the data and in the
averaged mock data sets. In the both panels, the thick line is the
mean dispersion in the data and the thin line is the mean dispersion
in the averaged mock data sets. The 1.0 $\mathrm{M}_{\oplus}$ case is
the top panel; the 0.5 $\mathrm{M}_{\oplus}$ case is the bottom panel.
In each panel, 90\% confidence bands are drawn about the mean
dispersion in the Hyades data (dotted lines) and the averaged mock
data sets (dashed lines). From Figure \ref{disp} it is clear that
neither the 0.5 nor the 1.0 $\mathrm{M}_{\oplus}$ case is ruled out by
the observed dispersion in the Hyades data at the 90\% confidence
level. However, looking at the trends apparent in these two cases, it
is clear that pollution at the mean level of $\sim$1.5
$\mathrm{M}_{\oplus}$ of iron is ruled out by the data.



\section{Conclusions}

Stars with giant planets in tight orbits, such as $\tau$ Boo, are likely
to have accreted $\sim$5 $\mathrm{M}_{\oplus}$ of iron \citep{mc}. A method is
presented for modifying standard stellar evolution models to
realistically account for this stellar pollution effects.  Such
polluted stars will exhibit a significantly hotter surface than stars
of similar mass and age with a correspondingly high bulk
metallicity. Polluted stellar evolution models may prove useful in
studying the evolution of stars which prove difficult for standard
stellar evolution models.
 
On the other hand, most stars in the solar neighborhood have not been
observed to have giant planets, and appear to have accreted
an average of 0.5 $\mathrm{M}_{\oplus}$ of iron \citep{mea}. At such a low
level, pollution effects in the CMD are found to be very small. As an
example, we look at pollution effects in the Hyades open
cluster. Stars in the Hyades all formed from the same material at
roughly the same time. At 650 Myr the stars we have chosen are all on
the main sequence and so past the era when pollution would have
occurred. Monte Carlo simulations show that the dispersion in $\Delta \mathrm{M_V}$ 
of the Hyades about a
polluted isochrone is consistent with the observed data,  assuming
that the pollution in the Hyades is at a similar level to the field
star sample. However, if the level of pollution in the Hyades were a factor 
of 3 or more
greater than that in the field stars, the analysis performed would be able to
rule it out. We are currently investigating the mass-[Fe/H] relationship in 
the Hyades to see if this data can put stronger constraints on stellar 
pollution.

\acknowledgments
Research supported in part by a NSF CAREER grant 0094231 to BCC.  BCC is a
Cottrell Scholar of the Research Corporation. We would like to thank Karl Gebhardt for guidance and computer programs relating to the LOWESS technique.

\end{document}